\documentclass{PoS}

\title{Analog hep-th, on Dirac materials and in general}

\ShortTitle{Analog hep-th, on Dirac materials and in general}

\author{\speaker{Alfredo Iorio}\\
        Institute of Particle and Nuclear Physics \\
        Faculty of Mathematics and Physics, Charles University \\
V Hole\v{s}ovi\v{c}k\'ach 2, 18000 Prague 8, Czech Republic
\\
        E-mail: \email{iorio@mff.cuni.cz}}

\abstract{The work of our group on reproducing scenarios of high energy theoretical physics on Dirac materials, like graphene, is illustrated. The main goal of this paper is to explain how versatile these systems are, and how far and wide into the hep-th territory we can explore with them. I first review why these materials lend themselves to the emergence of special relativistic-like matter and space, with the focus on the emergence of curvature. Then the crucial role of the low dimensions (2+1), and Weyl symmetry, towards the realization of a Unruh-kind of phenomenon (along with other interesting scenarios, that include the BTZ black hole and de Sitter spacetime) is explained. Comments on how far we went in the direction of experiments are offered too, followed by a list of some fresh results: From the time-loop to spot torsion, to the generalized uncertainty principle stemming from and underlying (lattice) length; From a model of grain-boundaries and their relation to (A)dS and Poincar\'{e} spacetime algebras, to Unconventional Supersymmetry and the role of the two Dirac points of graphene; and more. In the concluding remarks I briefly try to make the case for the realization of a ``CERN for analogs'', where theorists. both of the hep-th and of the cond-mat types, sit next to experimentalists, mostly of the cond-mat type.}

\FullConference{Corfu Summer Institute 2019 ``School and Workshops on Elementary Particle Physics and Gravity'' (CORFU2019)\\
		31 August - 25 September 2019\\ Corf\`{u}, Greece}

\def\be{\begin{equation}}
\def\ee{\end{equation}}

\def\bea{\begin{eqnarray}}
\def\eea{\end{eqnarray}}

\begin{document}

\section{Introduction}

The noble father of analogs is Richard Feynman, who, in a famous lecture titled ``Electrostatic Analogs'', available in \cite{Feynman} (see also the critical study offered in \cite{twostories}), says ``[...] The equations for many different physical situations have exactly the same appearance. Of course, the symbols may be different - one letter is substituted for another - but the mathematical form of the equations is the same. This means that having studied one subject, we immediately have a great deal of \textit{direct and precise knowledge} about the solutions of the equations of another''.  He then looks for a possible explanation of why analog systems do describe the same physics, which leads him to the thrilling hypothesis of \textit{more elementary constituents} than the ones we deem to be fundamental. Amazingly, these are also the conclusions of certain completely independent arguments of the contemporary research in QG, based on the Bekenstein maximal bound for the number of degrees of freedom of any physical system \cite{bekenstein}, see also \cite{bousso}. Amusingly, the names independently given to this fundamental level, both by Feynman and by Bekenstein, resemble each others: ``Xons'', and ``level X'', respectively.

Clearly, this elevates the epistemological status of analogs, from mere curiosities to reliable tests of fundamental theories. Nonetheless,despite Feynman's arguments are many decades old, this is not (yet) appreciated by the hep-th community. It is one of the side intents of this Project to amend that.

The more standard story that is usually told on the specific field of \textit{analog gravity}~\cite{Barcel2011}, starts with the seminal paper of Unruh of 1981, where he proposes to search for experimental signatures of his and Hawking effects, in a fluid dynamical analog~\cite{UnruhAnalog}. Due to our deeper understanding and experimental control of condensed matter systems, it is now becoming increasingly popular to reproduce that and other aspects of fundamental physics in analog systems. Examples include the Hawking phenomenon in Bose--Einstein condensates~ \cite{tris,Steinhauer:2015saa}, the Hawking/Unruh phenomenon on graphene~\cite{ioriolambiase1,ioriolambiase2}, gravitational and axial anomalies in Weyl semimetals~\cite{Gooth:2017mbd}, and more~\cite{Ulf_LeonhardtPRL2019}. Actually, gravity analogs are not limited to condensed-matter systems, as can be seen, e.g., by interpreting hadronization in heavy-ion collisions as a consequence of the Unruh effect~\cite{Castorina:2007eb}, (see also the recent \cite{shadowing}).

Despite those advances, there are still two milestones to reach. One is to understand the epistemic role of analogs in hep-th. The philosophical debate on this issue is still open (see, e.g.,~\cite{Dardashti2016,twostories}), and, as recalled earlier, not all theorists would agree that analogs are much more than mere divertissements. In fact, experimental results obtained in analogs are not used as feedbacks for the target theories they are analogs of.

A second milestone in the field is to eventually move to a dynamical set-up from a purely kinematical one (static background with an emergent horizon, where a quantum field experiences the wanted nontrivial phenomena), that is the current \textit{status quo} of the Hawking/Unruh phenomena on analogs. With this in our hands, we could eventually venture into the experimental study of various aspects of black-hole (BH) thermodynamics. Indeed, the interest is now turning to this type of analog experiments that could probe, e.g., the unitary nature of the BH evolution~\cite{PisinChen,carusottoWorkshop}. The great challenge there is a reliable definition of an analog BH entropy, or at least, of a QFT-like entanglement entropy that, in the presence of horizons, might serve the scope of setting-up some form of the second principle of BH thermodynamics.

Any progress in this direction would be truly important for the hep-th research. Having some results there, we could eventually be able to address the so-called {\it information paradox}, i.e., the apparent loss of information during BH evaporation, a question that, most probably, cannot be entirely solved via theoretical reasonings. In fact, some arguments state that in the presence of gravity, and especially at the extremal case of a BH, one has to expect some modifications of the quantum theory and, perhaps, deviations from unitary evolution~\cite{Penrose1996,Modak2015}. On the other hand, other arguments state that the evolution is always unitary and the information loss is prohibited~\cite{Thooft1990,Susskind1993,Hawking2004}. More recently, the impossibility to reconcile the unitary evolution, the principle of equivalence and the low energy effective QFT, gave rise to the controversial BH firewall phenomenon~\cite{Almheiri2013,Hooft2016}. This has further stimulated an upsurge of interest in scenarios trying to find a non-firewall resolutions of the above paradox~\cite{Hooft2016_I,Maldazena}. On the analog condensed matter side, theoretical work has shown over the years that BH physics can find an indirect realization in BEC systems~\cite{tris}, and thrilling experimental evidences have recently confirmed this fact~\cite{Steinhauer:2015saa}. The latter findings are often referred to as the first experimental examples of the Hawking effect.

For our work, there are two other important developments that are crucial. One is that, besides graphene itself, the family of Dirac materials (DMs) is now vastly populated, see, e.g.,~\cite{wehling}. The other is that the coherent-electron dynamics in graphene interacting with laser pulses, is a very hot topic in condensed-matter theory \cite{lasgraph1,lasgraph2}. As for the first point, the proposal of graphene as an analog of high-energy fundamental physics~\cite{ioriolambiase1,ioriolambiase2,iorio,pabloStran,ioriopaiswitten,grapheneQFTreview}, is based on its low-energy excitations~\cite{CastroNeto2009} being massless Dirac pseudo-relativistic fermions (the matter fields $\psi$), propagating in a carbon two-dimensional honeycomb lattice. The emergent (long-wave limit) description of the latter is a surface (piece of spacetime described by the ``emergent'' metric  $g_{\mu \nu}$). Such a behavior is shared by a wide range of materials, ranging from silicene and germanene through d-wave superconductors to topological insulators~\cite{wehling}. Each of those materials has its own peculiarities, which allow for
further extensions of results obtained with graphene, and hence permit to explore a wider range of the high-energy target systems. As for the second point, by learning from analogs based on laser-graphene interactions~\cite{lasgraph1,lasgraph2}, we shall be able to ascertain ways in which  nontrivial time-components of the $g_{\mu \nu}$ for the given DM should be constructed. 

In the following I illustrate the work of our group in Prague, by first recalling why DMs lend themselves as great testing beds of hep-th, and what is our point of view on that.

\section{The Weyl symmetry approach to curved graphene spacetimes}

Our research is presently largely devoted to the use of DMs as analog hep-th systems, suitable for the study of the responses of QFT to non-trivial geometries~\cite{iorio}, including the Hawking--Unruh phenomenon~\cite{ioriolambiase1,ioriolambiase2,pabloStran,ioriopaiswitten}. Reviews are available, such as \cite{grapheneQFTreview} and \cite{CastroNeto2009}, but we want to recall here the main ideas, and also point to our contribution (both recent and established) to the field.

\begin{figure}
\begin{center}
 \includegraphics[height= .3\textheight]{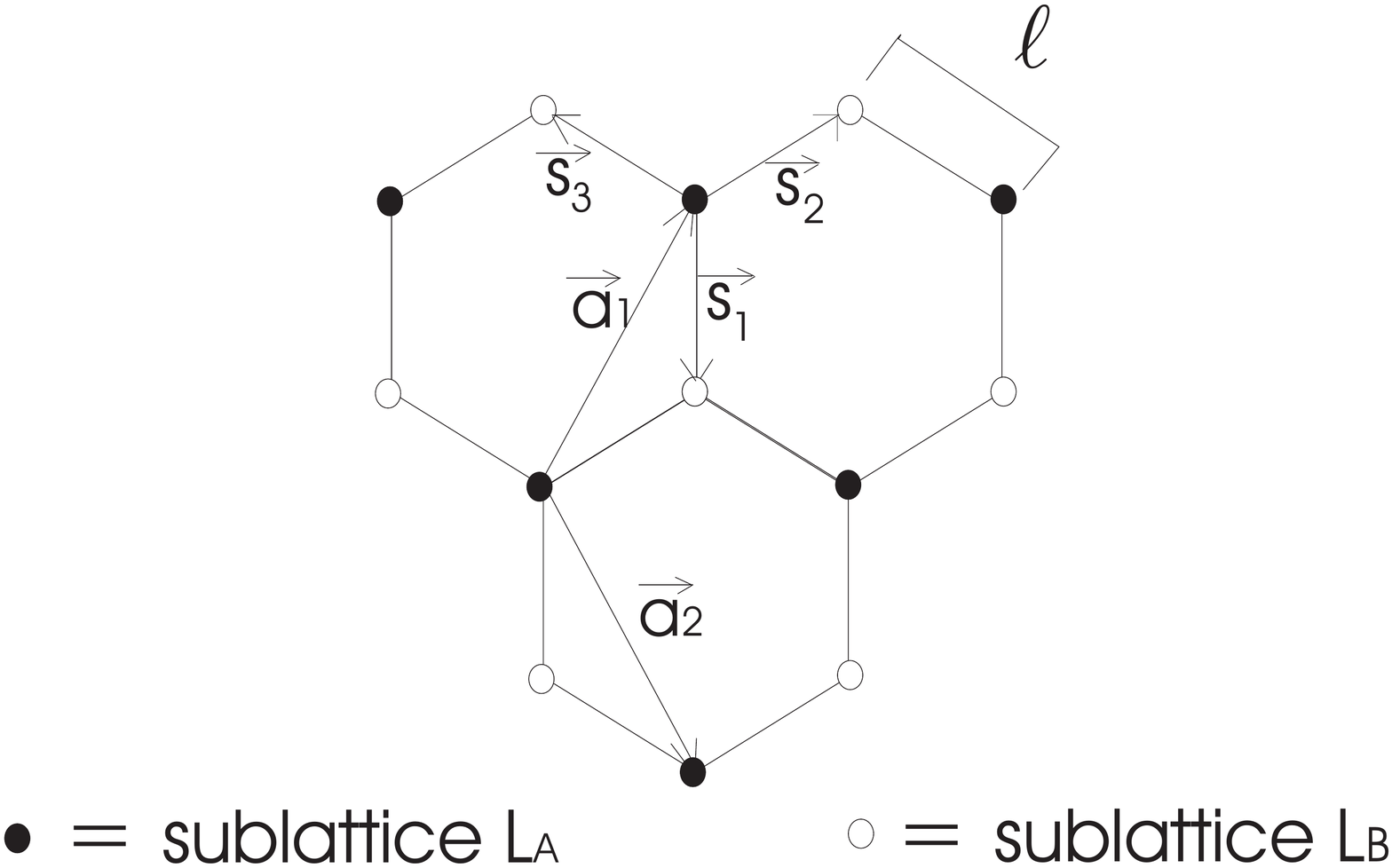}
\end{center}
  \caption{The honeycomb lattice of graphene, and its two triangular sublattices. The choice of the basis vectors, $(\vec{a}_1, \vec{a}_2)$ and
  $(\vec{s}_1, \vec{s}_2, \vec{s}_3)$, is, of course, not unique. Here we indicate the one used in \cite{grapheneQFTreview}.}
\label{honeycombpaper}
\end{figure}

Graphene is an allotrope of carbon. It is one-atom-thick, hence it is the closest in nature to a 2-dimensional object. It was first theoretically speculated about \cite{wallace}, and, decades later, experimentally found \cite{geimnovoselovFIRST}.
Its honeycomb lattice is made of two intertwined triangular sub-lattices, see Fig.~\ref{honeycombpaper}. As is by now well known, this structure is behind a natural description of its electronic properties in terms of massless, (2+1)-dimensional, Dirac (hence, relativistic-like) quasi-particles. Such electronic properties, in the tight-binding low-energy approximation, are customarily described by the Hamiltonian (we use $\hbar = 1$)
\begin{eqnarray}
    H = - \eta \sum_{\vec{r} \in L_A} \sum_{i =1}^3 \left( a^\dagger (\vec{r}) b(\vec{r} +\vec{s}_i)
    + b^\dagger (\vec{r} +\vec{s}_i) a (\vec{r}) \right) \label{primaH} \;,
\end{eqnarray}
where the nearest-neighbor hopping energy is $\eta \simeq 2.8$~eV, and $a, a^\dagger$ ($b, b^\dagger$) are the anti-commuting annihilation and creation operators, respectively, for the planar electrons in the sub-lattice $L_A$ ($L_B$), see Fig.~\ref{honeycombpaper}. All the vectors are bi-dimensional,  $\vec{r} = (x,y)$, and, for the choice of basis vectors made in Fig.~\ref{honeycombpaper}, if we Fourier transform, $a(\vec{r}) = \sum_{\vec{k}} a(\vec{k}) e^{i \vec{k} \cdot \vec{r}}$, etc,  then $H = \sum_{\vec{k}} ( f(\vec{k}) a^\dagger (\vec{k}) b(\vec{k}) + {\rm h.c.})$, with
\begin{equation}
  f(\vec{k}) = - \eta \; e^{-i \ell k_y} \left( 1 + 2 \; e^{i 3 \ell k_y / 2} \cos(\sqrt{3} \ell k_x / 2) \right) \;.
\end{equation}
Solving $E(\vec{k}) = \pm |f (\vec{k})| \equiv 0$ tells us where, in the first Brillouin zone (FBZ), conductivity and valence bands touch (if they do). Indeed, for graphene, this happens, pointing to a gapless spectrum, for which we expect massless excitations to emerge. Furthermore, the solution is not a Fermi \textit{line} (the $(2+1)$-dimensional version of the Fermi surface of the $(3+1)$ dimensions), but rather they are two Fermi \textit{points},
$ \vec{k}^D_\pm = \left( \pm \frac{4 \pi}{3 \sqrt{3} \ell}, 0 \right)$. There are actually six such points, but only the indicated two are inequivalent.

\begin{figure}
 \centering
  \includegraphics[height=.2\textheight]{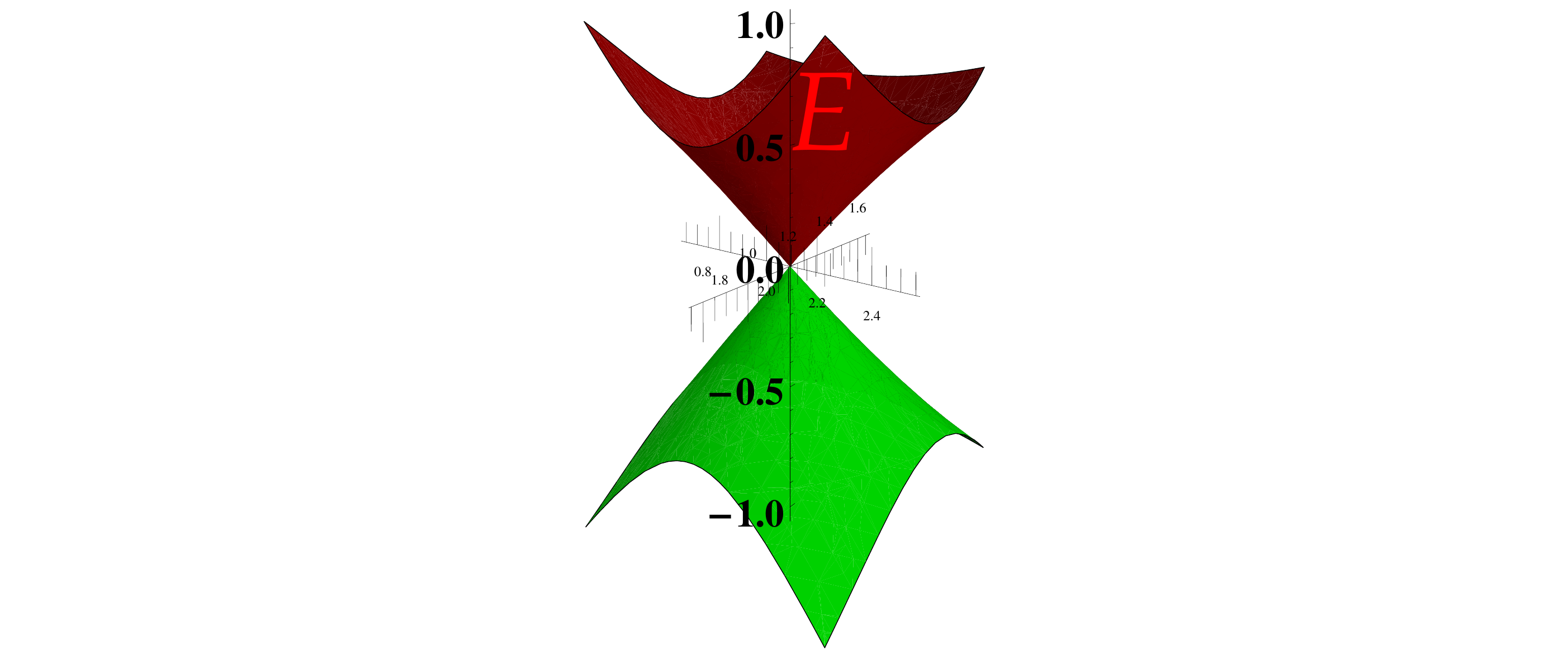}
  \caption{The linear dispersion relations near one of the Dirac points, showing the typical behavior of a relativistic-like system (the ``$v_F$-light-cone'' in $k$-space).}
\label{lindisprel}
\end{figure}

The label ``$D$'' on the Fermi points stands for ``Dirac''. That refers to the all-important fact that, near those points  the spectrum is {\it linear}, as can be seen from Fig.~\ref{lindisprel}, $E_{\pm} \simeq \pm v_F |\vec{k}|$, where $v_F = 3 \eta \ell / 2 \sim c/300$ is the Fermi velocity. This behavior is expected in a relativistic theory, whereas, in a non-relativistic system, the dispersion relations are usually quadratic.

If one linearizes around $\vec{k}^D_\pm$, $\vec{k}_\pm \simeq \vec{k}^D_\pm + \vec{p}$, then $f_+ (\vec{p}) \equiv f (\vec{k}_+) = v_F (p_x + i p_y)$,
$f_- (\vec{p}) \equiv f (\vec{k}_-) = - v_F (p_x - i p_y)$,  and $a_\pm (\vec{p}) \equiv a (\vec{k}_\pm)$,  $b_\pm (\vec{p}) \equiv b (\vec{k}_\pm)$, then the Hamiltonian (\ref{primaH}) becomes
\begin{eqnarray}
    H|_{\vec{k}_\pm} & \simeq & v_F \sum_{\vec{p}} \left(\psi_+^\dagger \vec{\sigma} \cdot \vec{p} \; \psi_+
    - \psi_-^\dagger \vec{\sigma}^* \cdot \vec{p} \; \psi_- \right) \label{hmeta}
\end{eqnarray}
where $\psi_\pm \equiv \left(\begin{array}{c} b_\pm \\ a_\pm \end{array}\right)$ are two-component Dirac spinors, and $\vec{\sigma} \equiv (\sigma_1, \sigma_2)$, $\vec{\sigma}^* \equiv (\sigma_1, - \sigma_2)$, with $\sigma_i$ the Pauli matrices.

Hence, if one considers the linear/relativistic-like regime only, the first scale is
\be
E_\ell \sim v_F / \ell \sim 4.2 \; {\rm eV}. \label{Escale}
\ee
Notice that $E_\ell \sim 1.5 \eta$, and that the associated wavelength, $\lambda = 2 \pi / |\vec{p}| \simeq 2 \pi v_F / E$, is $2 \pi \ell$. The electrons' wavelength, at energies below $E_\ell$, is large compared to the lattice length, $\lambda > 2 \pi \ell$. Those electrons see the graphene sheet as a continuum.

The two spinors are connected by the inversion of the full momentum $\vec{k}^D_+ + \vec{p} \to - \vec{k}^D_+ - \vec{p} \equiv \vec{k}^D_- - \vec{p}$. Whether one needs one or both such spinors to describe the physics, strongly depends on the given set-up. For instance, when only strain is present, one Dirac point is enough (see, e.g., \cite{pabloStran}), similarly (see below here) when certain approximations on the curvature are valid \cite{ioriolambiase1,ioriolambiase2}. The importance and relevance of the two Dirac points for emergent hep-th descriptions has been discussed at length in \cite{ioriopaiswitten}. There we have explored the role of grain boundaries and related necessity for two Dirac points, and their relation to the emergent spacetime torsion. The full focus on torsion, though, is in the recent \cite{tloop}.

When only one Dirac point is necessary, over the whole linear regime, the following Hamiltonian well captures the physics of undeformed (planar and unstrained) graphene
\begin{equation}
    H    = - i v_F \int d^2 x \; \psi^\dagger \vec{\sigma} \cdot \vec{\partial} \; \psi \;, \label{HGrapheneBpaper}
\end{equation}
where the two component spinor is, e.g., $\psi \equiv \psi_+$, we moved back to configuration space, $\vec{p} \to - i \vec{\partial}$, and sums turned into integrals because of the continuum limit. In various papers, we have exploited this regime to a great extent, till the inclusion of curvature and torsion in the geometric background. On the other hand, we also have investigated the regimes beyond the linear one, where granular effects associated to the lattice structure emerge, see \cite{GUP} and also the related \cite{GUPBTZ}.

When both Dirac points are necessary, one needs to consider four component spinors
$\Psi \equiv \left( \begin{array}{c} \psi_+ \\ \psi_- \\ \end{array} \right)$, and $4 \times 4$ Dirac matrices $\alpha^i = \left(\begin{array}{cc} \sigma^i & 0 \\ 0 & - {\sigma^*}^i \\ \end{array} \right)$, $\beta = \left(\begin{array}{cc} \sigma^3 & 0 \\ 0 & \sigma^3 \\ \end{array} \right)$, $i = 1, 2$. These matrices satisfy all the standard properties, see, e.g., \cite{ioriopaiswitten} and \cite{grapheneQFTreview}.

With these, the Hamiltonian is
\begin{equation}
H  =  - i v_F \int d^2 x \left( \psi_+^\dagger \vec{\sigma} \cdot \vec{\partial} \; \psi_+
    - \psi_-^\dagger \vec{\sigma}^* \cdot \vec{\partial} \; \psi_- \right) =
    - i v_F \int d^2 x \; \bar{\Psi} \vec{\gamma} \cdot \vec{\partial} \; \Psi \;.  \label{Hdiracgraphene1}
\end{equation}

In \cite{iorio} our goal was to identify the conditions for which graphene might get as close as possible to a full-power QFT in curved spacetime. Therefore, key issues had to be faced, such as the proper inclusion of the time variable in a relativistic-like description, and the role of the nontrivial vacua and their relation to different quantization schemes for different observers. All of this finds its synthesis in the Unruh or the Hawking effects, the clearest and unmistakable signatures of QFT in curved spacetime. Therefore, starting from \cite{iorio}, this road was pursued in \cite{ioriolambiase1,ioriolambiase2}. Let us explain here the main issues an the approximations made there.

Besides the scale (\ref{Escale}), when we introduce curvature, we also have a second scale. When this happens, $E_\ell$ is our ``high energy regime''. This is so because we ask the curvature to be small compared to a limiting maximal curvature, $1/\ell^2$, otherwise: i) it would make no sense to consider a smooth metric, and ii) $r < \ell$ (where $1/r^2$ measures the intrinsic curvature), means that we should bend the very strong $\sigma$-bonds, an instance that does not occur. Therefore, our second scale is
\be
E_r \sim v_F / r  \;,
\ee
with $E_r =  \ell / r \; E_\ell  < E_\ell$. To have a quantitative handle on these scales, let us take, e.g., $r \simeq 10 \ell$ as a small radius of curvature (high intrinsic curvature). To this corresponds an energy $E_r \sim 0.4$eV, whereas, to $r \sim 1 {\rm mm} \sim 10^6 \ell$, corresponds $E_r \sim 0.6 \mu$eV. The ``high energy'' to compare with is $E_\ell \sim 4$eV.

When energies are within $E_r$ (wavelengths comparable to $2 \pi r$) the electrons experience the global effects of curvature. That is to say that, at those wavelengths, they can distinguish between a flat and a curved surface, and between, e.g., a sphere and a pseudosphere. Therefore, whichever curvature $r < \ell$ we consider, the effects of curvature are felt until the wavelength becomes comparable to $2 \pi \ell$. The formalism we have used, though, takes into account all deformations of the geometric kind, with the exception of torsion. Hence, this includes intrinsic curvature, and elastic strain of the membrane (on the latter see \cite{pabloStran}), but our predicting power stops before $E_\ell$, because there local effects (such as the actual structure of the defects) play a role that must be taken into account into a (quantum gravity) QG type of theory. On the latter the first steps were moved in \cite{GUP} (see also the related \cite{GUPBTZ}).

The intrinsic curvature is taken here as produced by disclination defects, that are customarily described in elasticity theory (see, e.g., \cite{Kleinert}), by the (smooth) derivative of the (non-continuous) SO(2)-valued rotational angle $\partial_i {\omega} \equiv {\omega_i}$, where $i=1,2$ is a curved spatial index (see the Introduction for notation on indices etc). The corresponding (spatial) Riemann curvature tensor is easily obtained
\begin{equation}\label{a11}
    {R^{i j}}_{k l} =
    \epsilon^{i j} \epsilon_{k l} \epsilon^{m n} \partial_{m} \omega_{n} =
    \epsilon^{i j} \epsilon_{l k} 2 {\cal K}.
\end{equation}
where $\cal K$ is the Gaussian (intrinsic) curvature of the surface. In our approach we have included time, although the metric we adopted is
\begin{equation}\label{mainmetric}
g^{\rm graphene}_{\mu \nu}  = \left(\begin{array}{cc} 1 & 0  \quad 0 \\ \begin{array}{c} 0 \\ 0 \end{array} & - g_{i j} \\ \end{array} \right)\;,
\end{equation}
i.e., the curvature is all in the spatial part, and $\partial_t g_{i j}= 0$. Since the time dimension is included, the SO(2)-valued (abelian) disclination field has to be lifted-up to a SO(1,2)-valued (non-abelian) disclination field\footnote{Recall that in three dimensions $\omega_{\mu \; a b} = \epsilon_{a b c} \,\omega_\mu^{\;\; c}$.}, ${\omega_\mu}^a$, $a=0,1,2$, with $\omega_\mu^{\; a} = e^b_\mu \omega_b^{\; a}$ and the expression
\begin{equation}\label{omega3d}
\omega_a^{\; d}  = \frac{1}{2} \epsilon^{b c d} \left( e_{\mu a} \partial_b E_c^\mu + e_{\mu b} \partial_a E_c^\mu + e_{\mu c} \partial_b E_a^\mu \right) \,,  
\end{equation}
gives the relation between the disclination field and the metric (dreibein). All the information about intrinsic curvature does not change. For instance, the Riemann curvature tensor, ${R^\lambda}_{\mu \nu \rho}$, has only one independent component, proportional to $\cal K$, just like in (\ref{a11}) (see \cite{iorio}).

When only curvature is important, the long wavelength/small energy electronic properties of graphene, are well described by the following action
\begin{equation}\label{actionAcurvedpaper}
{\cal A} = i  v_F \int d^3 x \sqrt{g} \; \bar{\Psi} \gamma^\mu (\partial_\mu + \Omega_\mu) \Psi \;,
\end{equation}
with $\Omega_\mu \equiv {\omega_\mu}^a J_a$, and $J_a$ are the generators of SO(1,2), the local Lorentz transformations in this lower-dimensional setting. Notice that $J_a$ can never take into account mixing of the $\psi_\pm$,  because they are of the form $J^a = \left(\begin{array}{cc} j^a_+ & 0 \\ 0 & j^a_- \\ \end{array} \right)$, whereas, what is necessary are generators of the form $K^a = \left(\begin{array}{cc} 0 & k^a_+ \\  k^a_- & 0 \\ \end{array} \right)$. In \cite{ioriopaiswitten} we have discussed at length this point, within the Witten approach \cite{witten3dgravity} for which the most general gauge field, that takes into account strain, curvature (intrinsic and extrinsic) and torsion is of the form $A_\mu = \Omega_\mu + {\cal K}_\mu$, hence a Poincar\'{e} ($ISO(2,1)$) or (A)dS type of gauge theory, depending on the role played in here by the cosmological constant (on this see our work \cite{ioriolambiase1,ioriolambiase2}, and the review \cite{grapheneQFTreview}. The matter, though, might be faced by taking an alternative view, for which the gauge fields are internal rather than spatiotemporal. In this case, a link with the supersymmetry (SUSY) introduced by Zanelli and coworkers (that is a SUSY without superpartners, often referred to as unconventional SUSY (USUSY)) \cite{susyZanelli1} can be established, as we have shown in \cite{ioriopaiswitten}, and other authors in various papers, see \cite{susyZanelli2}, and especially the recent \cite{u-susy-graphene}.

Let us clarify here a point that is important for our future work. Within this approach, a nontrivial $g_{00}$ in (\ref{mainmetric}), hence a clean nontrivial general relativistic effect (recall that $g_{00} \sim V_{grav}$) can only happen if specific symmetries and set-ups map the lab system into the wanted one. Lot of work went into it, see, e.g, \cite{ioriolambiase1,ioriolambiase2,grapheneQFTreview}, and went as far as producing measurable predictions of a Hawking/Unruh effect, for certain specific shapes. Let us recall here the main ideas behind this approach (that we named the Weyl symmetry approach).

First of all, one notices that the action (\ref{actionAcurvedpaper}) enjoys local Weyl symmetry
\begin{equation}
g_{\mu \nu} \to e^{2 \sigma(x)} g_{\mu \nu} \, \quad {\rm and} \quad \Psi \to e^{-\sigma(x)} \Psi \;,
\end{equation}
that is an enormous symmetry among fields/spacetimes \cite{lor}. In most of our previously cited works on graphene, we focused on surfaces of constant $\cal K$. As explained in \cite{iorio}, to make the most of the Weyl symmetry of (\ref{actionAcurvedpaper}), we better focus on conformally flat metrics. The simplest metric to obtain in a laboratory is of the kind (\ref{mainmetric}). For this metric the Ricci tensor is ${R_\mu}^\nu = {\rm diag}(0, {\cal K}, {\cal K})$. This gives as the only nonzero components of the Cotton tensor, $C^{\mu \nu} = \left( \epsilon^{\mu \sigma \rho} \nabla_\sigma {R_\rho}^\nu + \mu \leftrightarrow \nu\right)$, the result $C^{0 x} = - \partial_y {\cal K} = C^{x 0} $ and $C^{0 y} = \partial_x {\cal K} = C^{y 0}$. Since conformal flatness in (2+1) dimensions amounts to $C^{\mu \nu} = 0$, this shows that all surfaces of constant $\cal K$ give raise in (\ref{mainmetric}) to conformally flat (2+1)-dimensional spacetimes. This points the light-spot to surfaces of constant Gaussian curvature.

The result $C^{\mu \nu} = 0$ is intrinsic (it is a tensorial equation, true in any frame), but to exploit Weyl symmetry to extract non-perturbative exact results, we need to find the coordinate frame, say it $Q^\mu \equiv (T,X,Y)$, where
\begin{equation}\label{genexplicitconfflat}
g^{\rm graphene}_{\mu \nu}  (Q) = \phi^2(Q) g^{\rm flat}_{\mu \nu} (Q) \;.
\end{equation}
Here, besides the technical problem of finding these coordinates (e.g., see the recent solution we have found for a specific interesting set-up \cite{kus}), the issue to solve is the physical meaning of the coordinates $Q^\mu$, and their practical feasibility.

Tightly related to the previous point is the issue of a conformal factor that makes the model {\it globally predictive, over the whole surface/spacetime}. The simplest possible solution would be a single-valued, and time independent $\phi(q)$, already in the original coordinates frame, $q^\mu \equiv (t,u,v)$, where $t$ is the laboratory time, and, e.g., $u, v$ the meridian and parallel coordinates of the surface.

Here we are dealing with a spacetime that is embedded into the flat (3+1)-dimensional Minkowski. Although, as said, we focused on intrinsic curvature effects, just like in a general relativistic context, issues related to the embedding, even just for the spatial part, are important. For instance, when the surface has negative curvature, we need to move from the abstract objects of non-Euclidean geometry, to objects measurable in a Euclidean real laboratory. This involves the last issue above about global predictability, and, in the case of negative curvature, necessarily leads to singular boundaries for the surfaces, as proved in a theorem by Hilbert, see, e.g., \cite{grapheneQFTreview} and \cite{icrystals, icrystals2}. Even the latter fact is, once more, a coordinates effect, due to our insisting in embedding in $\mathbb{R}^3$, and clarifies the hybrid nature of these emergent relativistic settings. We have then identified the quantum vacuum of the field that properly takes into account the measurements processes, as for any QFT on a curved spacetime, and how the graphene hybrid situation can realize that. As well known, this is crucial in QFT, in general, and on curved space, in particular.

\begin{figure}
 \centering
  \includegraphics[height=.3\textheight,angle =40]{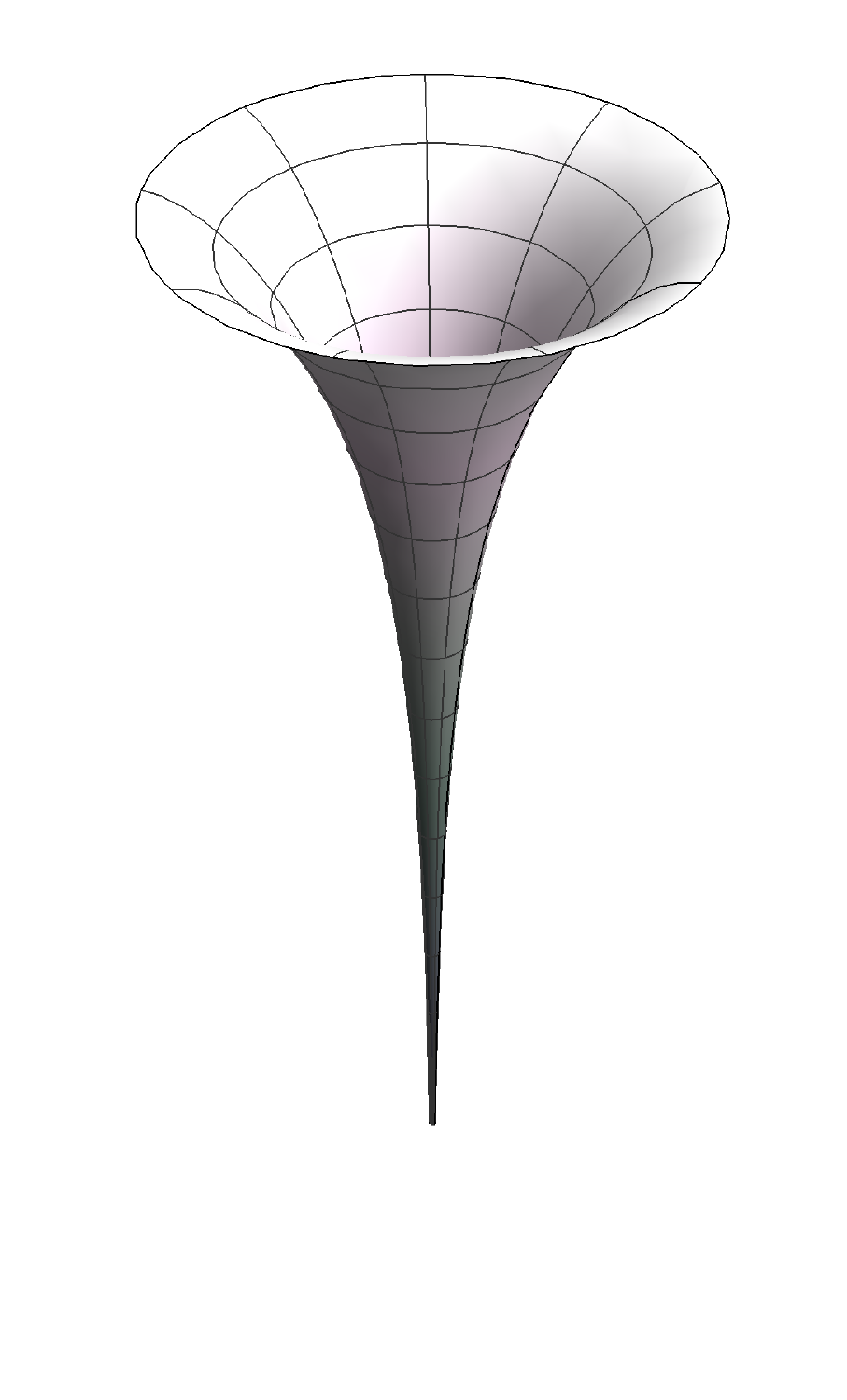}
  \caption{The most interesting three-dimensional spacetime metric we could engineer was obtained by acting only on the spatial part, shaped as a Beltrami pseudosphere.}
\label{defects}
\end{figure}

The above lead us to propose a variety of set-ups, the most promising being the one obtained by shaping graphene as shown in Fig. \ref{defects}, a configuration that we could punt into contact with three key spacetimes with horizon: the Rindler, the de Sitter and the BTZ BH \cite{BTZ1992}. The predicted impact on measurable quantities is reported in the first papers, and then explored in the subsequent efforts of computer-based simulations.

\section{Summary of our most recent results}

Let us now briefly summarize our most recent results, that have been encountered scattered here and there in the previous overview, or that we could not fit in there.

\subsection{Entropy of BHs}

In \cite{scholtz} (and in the forthcoming \cite{smaldone}) we provide general arguments regarding the connection between low-energy theories (gravity and quantum field theory) and a hypothetical fundamental theory of quantum gravity, under the assumptions of (i) validity of the holographic bound and (ii) preservation of unitary evolution at the level of the fundamental theory. In particular, the appeal to the holographic bound imposed on generic physical systems by the Bekenstein entropy implies that both classical geometry and quantum fields propagating in it should be regarded as phenomena emergent from the dynamics of the fundamental theory. The reshuffling of the fundamental degrees of freedom during the unitary evolution then leads to an entanglement between geometry and quantum fields. The consequences of such scenario are considered in the context of BH evaporation, see, e.g., \cite{Page1993a}, and the related information-loss issue: we provide a simplistic toy model in which an average loss of information is obtained as a consequence of the geometry-field entanglement. Pivotal for the previous study is the work of \cite{entaentro}, where we studied the Hawking--Unruh phenomenon in an entropy-operator approach that discloses the thermal properties of BH, from a new and instructive point of view.

\subsection{Generalized Uncertainty Principle on DMs}

In \cite{GUP} (see also the recent general results obtained in \cite{GUPBTZ}) we have studies the realization in DMs of specific GUPs associated to the existence of a fundamental length scale. There we showed that a generalized Dirac structure survives beyond the linear regime of the low-energy dispersion relations. A GUP of the kind compatible with specific quantum gravity scenarios with a fundamental minimal length (there the graphene lattice spacing) and Lorentz violation (there the particle/hole asymmetry, the trigonal warping, etc.) is naturally obtained. We then show that the corresponding emergent field theory is a table-top realization of such scenarios, by explicitly computing the third order Hamiltonian, and giving the general recipe for any order. Remarkably, our results imply that going beyond the low-energy approximation does not spoil the well-known correspondence with analogue massless quantum electrodynamics phenomena (as usually believed), but rather it is a way to obtain experimental signatures of quantum-gravity-like corrections to such phenomena.

\subsection{Grain boundaries on DMs and two hep-th scenarios: Witten 3d gravity, and USUSY}

In \cite{ioriopaiswitten} we proposed two different high-energy-theory correspondences with DMs associated to grain boundaries, that are topological defects for which both Dirac points are necessary. The first correspondence points to a $(3+1)$-dimensional theory, with nonzero torsion, with spatiotemporal gauge group $SO(3,1)$, locally isomorphic to the Lorentz group in $(3+1)$ dimensions, or to the de Sitter group in $(2+1)$ dimensions, in the spirit of 2+1-dimensional gravity \`{a}~la  Witten~\cite{witten3dgravity}. The other correspondence treats the two Dirac fields as an internal symmetry doublet, and it is linked there with Zanelli's USUSY~\cite{susyZanelli1} with $SU(2)$ internal symmetry. Our results pave the way to the inclusion of grain boundaries in the emergent field theory picture associated with these materials, whereas disclinations and dislocations have been already well explored.

\subsection{Particle-Hole pairs in graphene to spot spatiotemporal torsion}

In \cite{tloop}, assuming that dislocations could be meaningfully described by torsion, we proposed a scenario based on the role of time in the low-energy regime of two-dimensional Dirac materials, for which coupling of the fully antisymmetric component of the torsion with the emergent spinor is not necessarily zero. Our approach is based on the realization of an exotic \textit{time-loop}, that could be seen as oscillating particle-hole pairs, an instance that was later considered in \cite{prlpair}. Although this is a theoretical paper, we moved there tje first steps toward testing the realization of these scenarios, by envisaging \textit{Gedankenexperiments} on the interplay between an external electromagnetic field (to excite the pair particle-hole and realize the time-loops), and a suitable distribution of dislocations described as torsion (responsible for the measurable holonomy in the time-loop, hence a current). Our general analysis there establishes that we need to move to a nonlinear response regime. We then conclude by pointing to recent results from the interaction laser-graphene that could be used to look for manifestations of the torsion-induced holonomy of the time-loop, e.g., as specific patterns of suppression/generation of higher harmonics.

\subsection{Vortex solutions of Liouville equation and quasi spherical surfaces}

In \cite{kus} we identified the two-dimensional surfaces corresponding to certain solutions of the Liouville equation of importance for mathematical physics, the non-topological Chern-Simons (or Jackiw-Pi \cite{jackiwpi}) vortex solutions, characterized by an integer\cite{Horvathy_Yera} $N \ge 1$. Such surfaces, that we called there $S^2 (N)$, have positive constant Gaussian curvature, $K$, but are spheres only when $N=1$. They have edges, and, for any fixed $K$, have maximal radius $c$ that we find here to be $c = N / \sqrt{K} $. If such surfaces are constructed in a laboratory by using DMs, our findings could be of interest to realize table-top Dirac massless excitations on nontrivial backgrounds. We also briefly discuss the type of three-dimensional spacetimes obtained as the product $S^2 (N) \times \mathbb{R}$.

\subsection{BH Entropy and QCD}

Although this paper is entirely focussed on DM as analog systems, let us here briefly mention another direction that our group is pursuing in a similar spirit, that is to probe how close the physics of BHs is to Quantum Chromo-Dynamics (QCD).

The interpretation of quark confinement as the effect of a classical event horizon for color degrees of freedom, naturally lead to view hadronization as the quantum tunnelling
through such horizon \cite{Castorina:2007eb}. From this point of view, hadron formation is the result of the Unruh radiation associated to the strong force. More precisely, hadronization is the result of a Unruh phenomenon related with the string breaking/formation mechanism, that is, with the large distances QCD behavior. Within that picture in \cite{qcdentropy} we show that QCD entropy, evaluated by lattice simulations in the region $T_c < T < 1.3T_c$, is in reasonable agreement with a melting color event horizon.

Relying on the results of \cite{dima1}, that link entanglement entropy and parton distribution functions in deep inelastic scattering, and focusing on the small Bjorken scaling region in \cite{shadowing} we present indications that  gluon shadowing might indeed be explained as due to a depletion of the entanglement entropy, between observed and unobserved degrees of freedom, per nucleon within a nucleus, with respect to the free nucleon. We apply to gluon shadowing the general Page approach to the calculation of the entanglement entropy of an evaporating BH \cite{Page1993a}, giving physical motivations of the results.

\subsection{Realization in labs}

Besides that theoretical work, we always aimed at the actual realization of the necessary structures in real laboratories. See, e.g., the work \cite{icrystals, icrystals2}, where we realized Lobachevsky geometry via simulations, by producing a carbon-based mechanically stable molecular structure, arranged in the shape of a Beltrami pseudosphere. We found there that that structure: i) corresponds to a non-Euclidean crystallographic group, namely a loxodromic subgroup of $SL(2,\mathbb{Z})$; ii) has an unavoidable singular boundary, that we fully take into account. Our approach, substantiated by extensive numerical simulations of Beltrami pseudospheres of different size, might be applied to other surfaces of constant negative Gaussian curvature, and points to a general procedure to generate them. Our results there pave to future experiments.

\section{Concluding remarks on the need for HELIOS}

Let me close by making the case for a laboratory where fundamental theories of nature are tested with analogs. What I refer to here is a ``CERN for analogs'', i.e., a facility where high-energy theorists, of all kinds (cosmologists, string theories, quantum gravity experts, etc), as well as condensed matter theorists, sit next to dedicated analog experiments that run to test their theories. To name things is a crucial part of making them happens. That is why, it is a while that within our group we call this facility HELIOS, see, e.g., \cite{reach the unreachable,reach the unreachable2,reach the unreachable3} an evocative Greek name for something that should shed light on the darkness of the unknown, and an acronym for ``High Energy Laboratory for Indirect ObservationS''.

There are, at least, three reasons why the scientific community must make an effort towards the construction of HELIOS. First, we should be looking for the elementary constituents of matter (and spacetime) \textit{here and now}, starting from answering Feynman's deep question on why analogs do describe the same physics. In other words (Feynman's) we should look for \textit{Xons}, here and now. Second, clearly certain theories can never be tested directly, i.e., within the scenarios where they are supposed to be at work, henceforth we should try to test them indirectly. Third, being necessary to push the condensed matter systems to conditions that are unusual or even extreme for the standard agenda, the chance to unexpected spin-offs is clearly quite big.

\section*{Acnowledgements}

I happily thank Fedele Lizzi, George Zoupanos and all the organizers for the kind invitation. I am indebted to all the colleagues that made this collaborative work possible. I like to mention the late Martin Scholtz, in particular, for his intellectual courage, deep love of the elegance of physics, and his great heart. He is sorely missed. This work is partially supported by the grant UNCE/SCI/013.

\end{document}